**Zenon's Demon and the Denial of Domain-Generality for Transformer-Based Computational Models of Human Behavior**

[1†]Orr, M., [1]Cranford, E. A., [1]Ford, K., [1]Gluck, K., [1]Hancock, W., [2]Lebiere, C. [1]Pirolli, P., [3]Ritter, F. E., [4]Stocco, A.

[1]Florida Institute for Human and Machine Cognition, Pensacola, FL, USA
[2]Carnegie Mellon University, Pittsburgh, PA, USA
[3]Pennsylvania State University, State College, PA, USA
[4]University of Washington, Seattle, WA, USA
[†]morr@ihmc.org

**Transformer-based models of human behavior (e.g., the Centaur model by Binz, et al., 2025) posit to be domain general computational models of human behavior. The claim of domain-generality is by virtue of the supposed capability to predict and simulate human behavior across a vast range of cognitive and perceptual domains. Further, it is argued that this degree of performance places such models on a path toward general, unified theories of cognition (Newell, 1990). We contest this characterization. We propose the Domain-Generality Thesis: A computational model is domain-general if and only if it performs well across a structurally distinct set of tasks. While transformer-based models of human behavior achieve impressive statistical breadth, we demonstrate that, by example, they fail this structural criterion, conflating parametric variations of a single task with genuine cognitive diversity. W*e construct an argument that denies the domain-generality of transformer-based models of human behavior and thus denies the purported status as a start on the path towards general, unified theories of cognition.***

Foundational to this analysis is the rigorous definition of a Task Description within the paradigm of unified theories of cognition (Newell, 1990). We define this as the complete set of predicates governing the interaction between a simulation agent and its environment, encompassing the explicit instructions, the agent's goals, and—crucially—the specific algorithmic procedure used by the agent to execute the task. This definition moves beyond surface-level input variations to capture the structural components necessary for establishing "strong equivalence" (Pylyshyn, 1984). It is against this formal backdrop that the distinctness of Centaur's domain-generality must be judged.

To aid our analysis, we introduce Zenon's demon[1], a being who is infallible in the generation of Task Descriptions for any given experiment and in the comparison of any two experiments via the Task Descriptions[2]—it will, after comparison, report a truth value of TRUE if both task descriptions are identical, else it reports FALSE. The demon relies, critically, on the Task Description as specified above.

Let us conduct a thought experiment using the drifting four-armed bandit paradigm in which an agent attempts to maximize its reward across a series of experimental trials. Each trial provides the agent with four options; upon selection of any option a reward is given. The distribution of rewards across options is unknown to the subject and stochastic across trials.

Imagine that Zenon's demon builds a representation of the Task Description of the four-armed bandit experiment with a transformer-based model of human behavior. We will use Centaur as a representative agent case (Binz, et al, 2025). (We illustrate the demon's view in Figure 1.) The natural language expression of the experiment (as required by Centaur[3]) constitutes the environment of the agent and amounts to an ordered set of tokens. The demon would unpack the agent's (Centaur) procedure and objective for a single trial in the experiment: produce a single token given a set of ordered tokens as input in its (Centaur's) context window. The agent takes the environment (the set of all ordered tokens) as input into its context window and processes it via a forward pass through its neural network culminating in a distribution across all its possible tokens from which it returns, as output to the task environment, the most probable token. Then, the environment takes the agent's last output as input and appends it to the end of the ordered set of tokens, thus setting up a new environment for the agent's next input. The demon would realize that one run of an experiment, across multiple trials, constitutes a dialog between the agent and environment that iterates until the agent reaches some stopping criterion.

---

[1] In recognition that Zenon Pylyshyn's elucidation of strong equivalence in cognitive science spawned the demon.

[2] This includes actual human performed experiments; Zenon's demon can abstract the program used by a human for any given task.

[3] Because Centaur is the agent in this example, we must explain a central but subtle constraint in Centaur's method: it works if and only if there exists a suitable procedure for expressing the original experiment designed for and conducted with humans—its experimental paradigm and generated behavioral data—into natural language alone. This amounts to a short description of the task instructions in natural language followed by a natural language translation of behavior on each trial for a given human subject, e.g. (four-armed bandit): "You press <<L>> and get 84.0 points" is a string given for a single trial.

To make a comparison between the four-armed bandit and any other experiment in Centaur's repertoire, the demon would generate a new Task Description and compare it to the four-armed bandit experiment. How would Zenon's demon judge the comparison of any other Centaur experiment to the four-armed bandit Centaur experiment? Would it judge the two Task Descriptions as identical or different?

Zenon's demon would judge any two such Task Descriptions as identical—same task environment, same agent procedure/program, i.e., same Task Description. This would be so for any of the 12720 possible pair-wise comparisons across the 160 experiments in Centaur's repertoire. All experiments would be deemed the same. Variation in Task Descriptions across experiments would be zero. Domain-generality would be, thus, false.

This conclusion may seem, at first blush, incredulous or confusing. Revisiting Figure 1, Panel B, however, will highlight that the only difference between any two Centaur experiments is that initialization $\boldsymbol{I}$ in the task environment $\boldsymbol{E}$ will be a different set of ordered tokens, a trivial difference in terms of the structure of Task Descriptions.

If we asked our demon to explain its reasoning in plain English, we might hear something like this: "Well, the Centaur agent (a transformer), in essence, was tasked with reading the beginning of a story and then asked to finish the story. The story, it turns out, was about another agent, maybe a human, in some kind of psychological experiment, so the Centaur agent was trying to expand and complete the story about this other agent. For all the experiments I analyzed, this was the task of the Centaur agent. So, I deemed them all equivalent."

Zenon's demon sees clearly what is difficult for humans to see: the Centaur agent isn't part of the psychological experiment described but is simply reading about it and trying to predict how the story unfolds. Trivially, then, the demon concludes that the Task Descriptions are the same across experiments while the stories change.

Notice that it is not paradoxical to both deny Centaur's domain-generality and to celebrate its ability to finish (to predict) stories that are about human behavior in psychological experiments. Centaur, as an example of transformer-based models of human behavior, may be on a road towards increasingly accurate and contextual statistical prediction of some patterns of human behavior in psychological experiments. But, critically, this only applies to patterns of human behavior that are invariant to the transformation from experimental paradigms and associated behavioral data into natural language data. This, we assert, is not on the same path that leads towards domain-general, unified theories of cognition.

**A.**

```
Step 0: A(E(I )) → E(O_0)
Step 1: A(E(I,O_0)) → E(O_1)
Step 2: A(E(I,O_0,O_1)) → E(O_2)
Step 3: A(E(I,O_0,O_1,O_2)) → E(O_3)
...
...
...
Step F: A(E(I,O_0,O_1,O_2,...O_{F-3},O_{F-2},O_{F-1})) → E(O_F)
```

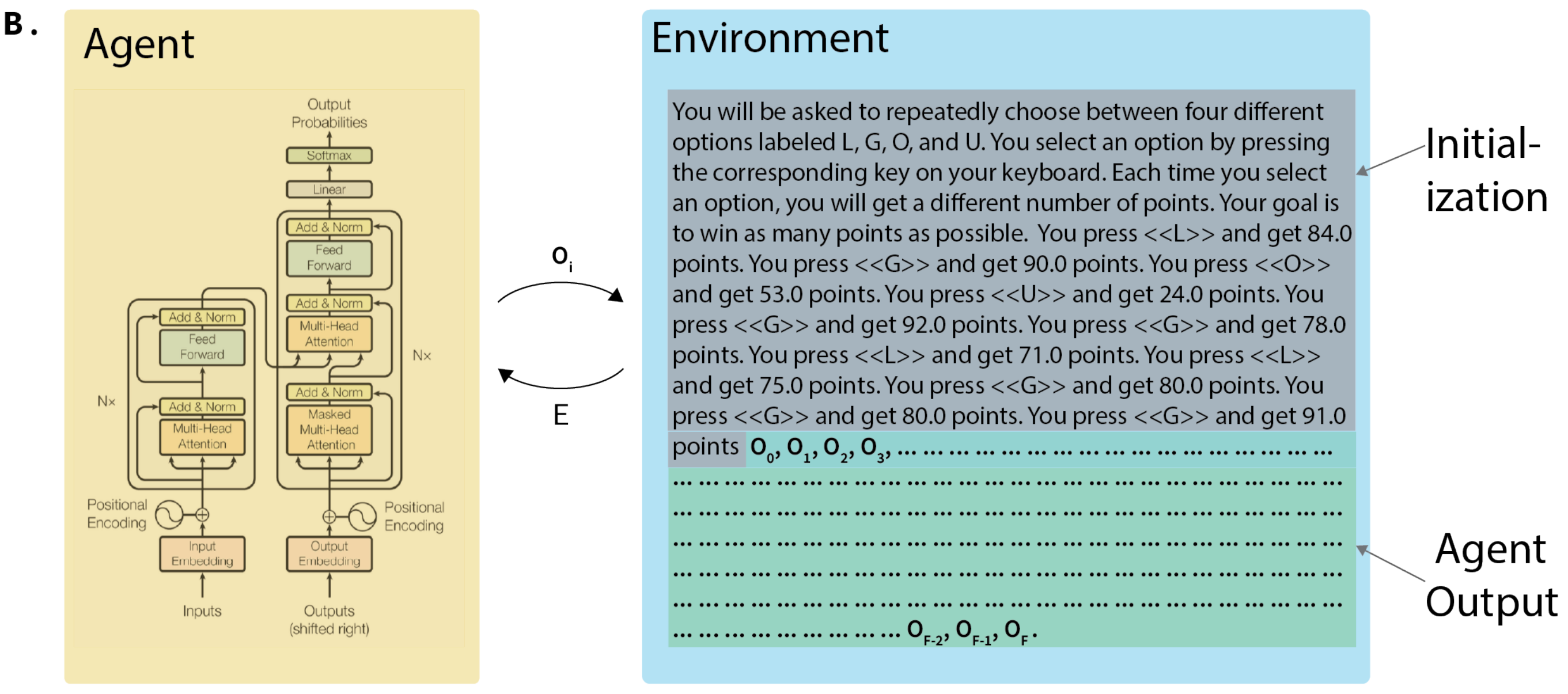

**Figure 1.** *The View for Zenon's Demon.* **Panel A:** A Centaur experiment is defined by the tuple (***A***, ***I***, ***E***), where ***E*** is the experimental environment (a set of ordered tokens), ***I*** is the initialized version of ***E***, and ***A*** is the Centaur agent. For any Centaur experiment, ***I*** is defined as the portion of the natural language translation of any given experiment from the first token of its natural language translation beyond, to some degree, the portion of the natural language translation that contains what are considered the instructions of the experiment. ***E*** holds ***I*** and will append to ***I*** the output ***O*** of ***A*** at any increment step ***i*** up until the final step ***F*** in the experiment. **Panel B:** A graphic depiction of Zenon's Demon's Task Description for the Centaur version of the Four-Armed Bandit Problem.

**Acknowledgements**
This work was supported by the US Defense Advanced Research Project Agency (DARPA) contract # HR00112590037.

**Author Contributions**
MGO wrote the initial draft of the manuscript. All other authors contributed on subsequent revisions of the original draft in terms of content, organization, and scientific advisement.

**Competing Interest Declaration**
We have no competing interests to declare in relation to this work.